\documentclass[aps,prl,showpacs,twocolumn]{revtex4}
\usepackage{amsmath,bm}
\usepackage[dvips]{graphicx}

\begin{document}

\title{Extreme Outages due to Polarization Mode Dispersion
}

\author{Vladimir Chernyak$^a$, Michael Chertkov$^b$, Igor
Kolokolov$^{b,c,d}$, and Vladimir Lebedev$^{b,c}$}

\affiliation{$^a$Corning Inc., SP-DV-02-8, Corning, NY 14831, USA; \\
$^b$Theoretical Division, LANL, Los Alamos, NM 87545, USA; \\
$^c$Landau Institute for Theoretical Physics, Moscow, Kosygina 2, 117334, Russia; \\
$^d$Budker Institute of Nuclear Physics, Novosibirsk 630090, Russia.}

\date{\today}

\begin{abstract}

  We investigate the dependence of the bit-error-rate (BER) caused by amplifier noise in
  a linear optical fiber telecommunication system on the fiber birefringence. We show that the
  probability distribution function (PDF) of BER obtained by averaging over many realizations
  of birefringent disorder has an extended tail corresponding to anomalously large values
  of BER. We specifically discuss the dependence of the tail on such details of the pulse
  detection at the fiber output as ``setting the clock" and filtering procedures.

\end{abstract}

\pacs{42.81.Gs, 78.55.Qr, 05.40.-a}

\maketitle

Transmission errors in modern optical telecommunication systems are caused by various impairments
(limiting factors). In systems with the transmisson rate 40 $Gb/s$ or higher, polarization mode
dispersion (PMD) is one of the major impairments. PMD leads to splitting and broadening an
initially compact pulse \cite{78RU,81MSK,87BPW,87ACMD}. The effect is usually characterized by the
so-called PMD vector that determines the leading PMD-related pulse distortion
\cite{88Pol,88PBWS,91PWN}. It is also recognized that the polarization vector does not provide a
complete description of the PMD phenomenon and some proposals aiming to account for
``higher-order" PMD effects have been recently discussed \cite{98Bul,00KNGJ,00ELMYT,02BKM}.
Birefringent disorder is frozen (i.e. it does not vary at least on the time scales corresponding
to the optical signal propagation). Optical noise originating from amplified spontaneous emission
constitutes an impairment of a different nature: The amplifier noise is short correlated on the
time scale of the signal width. In this letter we discuss the joint effect of the amplifier noise
and birefringent disorder on the BER. Our main goal is estimating the probability of special rare
configurations of the fiber birefringence that produce an anomalously large values of BER, and
thus determine the information transmission reliability. Evaluation of the signal BER due to the
amplifier noise for a given realization of birefringent disorder is the first step of our
theoretical analysis. Second, we intend to study the PDF (normalized histogram) of BER, where the
statistics is collected over different fibers or over the states of a given fiber at different
times, and focus on the probability of anomalously large BER. We analyze the basic (no
compensation) situation  and compare it with the case of the simplest compensation scheme known as
``setting the clock". More sophisticated compensation strategies will be discussed elsewhere.

The envelope of the optical field propagating in a given channel in the linear regime (i.e. at
relatively low optical power), which is subject to PMD distortion and amplifier noise, satisfies
the following equation \cite{79US,81Kam,89Agr}
 \begin{eqnarray}
 \partial_z{\bm\varPsi}-i\hat{\varDelta}(z){\bm\varPsi}
 -\hat{m}(z)\partial_t{\bm\varPsi}
 -id(z)\partial_t^2{\bm\varPsi}={\bm\xi}(z,t).
 \label{PMD} \end{eqnarray}
Here $z$, $t$, $\bm\xi$, and $d$ are the position along the fiber, retarded time, the amplifier
noise, and the chromatic dispersion, respectively. The envelope $\bm\varPsi$ is a two-component
complex field, the two components representing two states of the optical signal polarization. The
birefringent disorder is characterized by two random $2\times2$ traceless matrix fields related to
the zero-, $\hat{\Delta}$, and first-, $\hat{m}$, orders in the frequency expansion with respect
to the deviation from the carrier frequency $\omega_0$. Birefringence that affects the light
polarization is practically frozen ($t$-independent) on all propagation-related time scales. The
matrix $\hat{\varDelta}$ can be excluded from the consideration by the transformation
$\bm{\varPsi}\to\hat{V}\bm{\varPsi}$, $\bm{\xi}\to\hat{V}\bm{\xi}$ and
$\hat{m}\to\hat{V}\hat{m}\hat{V}^{-1}$. Here, the unitary matrix
$\hat{V}(z)=T\exp[i\int_0^z\mathrm d z'\hat{\varDelta}(z')]$ is the ordered exponential defined as
a formal solution of the equation $\partial_z\hat{V}=i\hat{\varDelta}\hat{V}$ with
$\hat{V}(0)=\hat{1}$. Hereafter, we will always use the renormalized quantities. We further
represent the solution of Eq. (\ref{PMD}) as $\bm\varPsi=\bm\varphi+\bm\phi$ where,
 \begin{eqnarray}
 \bm\varphi=\hat W(z)\bm\varPsi_0(t), \quad
 \bm\phi=\int_0^z\!\!\!\mathrm d z'\,
 \hat W(z) \hat W^{-1}(z') \bm\xi(z',t),
 \label{PMD4} \\
 \hat W(z)=\exp\left[i
 \!\!\int_0^z\!\!\!\mathrm dz'
 d(z')\partial_t^2\right]\!
 T\!\exp\left[\int_0^z\!\!\!\mathrm dz'
 \hat m(z')\partial_t\right]\!,
 \label{PDM3} \end{eqnarray}
and $\bm\varPsi_0(t)$ stands for the initial pulse shape.

We consider a situation when the pulse propagation distance substantially exceeds the
inter-amplifier separation (the system consists of a large number of spans). Our approach allows
to treat discrete (erbium) and distributed (Raman) amplification schemes within the same framework.
The additive noise, $\bm\xi$ generated by optical amplifiers is zero in average. The statistics of
$\bm\xi$ is Gaussian with spectral properties determined solely by the steady state features of
amplifiers (gain and noise figure) \cite{94Des}. The noise correlation time is much shorter than
the pulse temporal width, and therefore $\bm\xi$ can be treated as $\delta$-correlated in time.
Eqs. (\ref{PMD4},\ref{PDM3}) imply that the noise contribution to the output signal $\bm\phi$ is a
zero mean Gaussian field characterized by the following pair correlation function
 \begin{equation}
 \langle\phi_\alpha(Z,t_1)\phi^\ast_\beta(Z,t_2)\rangle
 =D_\xi Z\delta_{\alpha\beta} \delta(t_1-t_2).
 \label{phiphi} \end{equation}
and, therefore, is statistically independent of both $d(z)$ and $\hat{m}(z)$. Here, $Z$ is the total
system length, the product $D_\xi Z$ being the amplified spontaneous emission (ASE) spectral
density of the line. The coefficient $D_\xi$ is introduced into Eq. (\ref{phiphi}) to reveal the
linear growth of the ASE factor with $Z$ \cite{94Des}.

The matrix of birefringence $\hat{m}$ can be parameterized by a three-component real field $h_j$,
$\hat{m}=\sum h_j\hat{\sigma}_j$, with $\hat{\sigma}_j$ being a set of three Pauli matrices. The
field $\bm h$ is zero in average and short-correlated in $z$. The above transformation $\hat{m}\to
\hat{V}\hat{m}\hat{V}^{-1}$ guarantees the statistics of $h_j$ to be isotropic. Since $\bm h$
enters the observables described by Eqs. (\ref{PMD4},\ref{PDM3}) in an integral form the central
limit theorem (see, e.g., \cite{Feller}) implies that the field $h_j$ can be treated as a Gaussian
field with
 \begin{equation}
 \langle h_i(z_1)h_j(z_2)\rangle
 =D_m\delta_{ij}\delta(z_1-z_2),
 \label{hh} \end{equation}
and the average in Eq. (\ref{hh}) is taken over the birefringent disorder realizations
(corresponding to different fibers or states of a single fiber at different times). In case of
weak birefringent disorder the integral $\bm H=\int_0^Z\mathrm dz\,\bm h(z)$ represents the PMD
vector. Thus, $D_m=k^2/12$, where $k$ is the so-called PMD coefficient.

We consider the return-to-zero (RZ) modulation format when the pulses are well separated in $t$.
The signal detection at the line output, $z=Z$, corresponds to measuring the output pulse
intensity, $I$,
 \begin{equation}
 I=\int \mathrm dt\, G(t)
 \left|{\cal K}\bm\varphi(Z,t)
 +{\cal K}\bm\phi(Z,t)\right|^2 \,,
 \label{nnn} \end{equation}
where $G(t)$ is a convolution of the electrical (current) filter function with the sampling window
function. The linear operator ${\cal K}$ in Eq. (\ref{nnn}) stands for an optical filter and a
variety of engineering ``tricks'' applied to the output signal, $\bm\varPsi(Z,t)$. Out of the
variety of the ``tricks" we will discuss here only those correspondent to optical filtering and
``setting the clock" compensation. The latter can be formalized as, ${\cal K}_{cl}{\bm\varPsi}
={\bm\varPsi} (t-t_{cl})$, where $t_{cl}$ is an optimal time delay. Ideally, $I$ takes two
distinct values corresponding to the bits ``0'' and ``1'', respectively. However, the impairments
enforce deviations of $I$ from the ideal values. The output signal is detected by introducing a
threshold (decision level), $I_d$, and declaring that the signal codes ``1" if $I>I_d$ and ``0"
otherwise. Sometimes the information is lost, i.e. an initial ``1" is detected as ``0" at the
output or vise versa. The BER is the rate of such events which is extracted from measurement of
many pulses coming through a fiber with a given realization of the PMD disorder, $h_j(z)$. For
successful system performance the BER should be extremely small, i.e. typically both impairments
can cause only a small distortion of a pulse or, stated differently, the optical signal-to-noise
ratio (OSNR) and the ratio of the squared pulse width to the mean squared value of the PMD vector
are both large. OSRN can be estimated as $I_0/(D_\xi Z)$ where
$I_0=\int\mathrm dt\,|\varPsi_0(t)|^2$ is the initial pulse intensity and the integration goes 
over a single slot populated by an ideal (initial) pulse, coding ``1''.

Based on Eq. (\ref{nnn}) one concludes that the input ``0" is converted into the output ``1''
primarily due to the noise-induced contribution $\bm\phi$ and, therefore, the probability of such
event is insensitive to the PMD disorder in accordance with Eq. (\ref{phiphi}). Therefore,
anomalously large values of BER originate solely from the ``$1\to0$" transitions. Let $B$ be the
probability of such an event. Since OSNR is large, $B$ can be estimated using Eqs.
(\ref{phiphi},\ref{nnn}) as the probability of an optimal fluctuation of $\bm\phi$ leading to
$I=I_d$. Then one concludes that the product $D_\xi Z\ln B$ depends on the disorder, the chromatic
dispersion coefficient and the measurement procedure, while being insensitive to the noise
characteristics. Since OSNR is large, even a weak PMD disorder could produce a large increase in
the value of $B$. This fact allows a perturbative evaluation of the $B$-dependence on the
disorder, starting with an expansion of the ordered exponential (\ref{PDM3}) in powers of $\bm h$.
If no compensation is applied the linear term is prevailing. It is convenient to introduce a
dimensionless coefficient $\mu_1$ in accordance with $(D_\xi Z/I_0)\ln(B/B_0)=\mu_1H_3/b+O(H^2)$,
where the initial pulse $\bm\varPsi_0$ is assumed to be linearly polarized,
$\bm\varPsi_0$ ($b$ is the signal width), $B_0$ being a typical value of $B$ that
corresponds to $h_j=0$, $\ln B_0\sim -I_0/(D_\xi Z)$. ``Setting the clock" compensation cancels
out the linear in $H$ contribution, i.e. $\mu_1=0$ if $t_{cl}$ is chosen to be equal to $H_3$. In
this case and also when the output signal is not chirped (this e.g. corresponds to the case when
the initial signal is not chirped and the integral chromatic dispersion, $\int_0^Z\mathrm dz\,d$,
is negligible) one gets $(D_\xi Z/I_0)\ln(B/B_0) =\mu_2(H_1^2+H_2^2)/b^2+O(H^3)$, where $\mu_2$ is
another dimensionless coefficient.

\begin{figure}
 \includegraphics[width=0.4\textwidth]{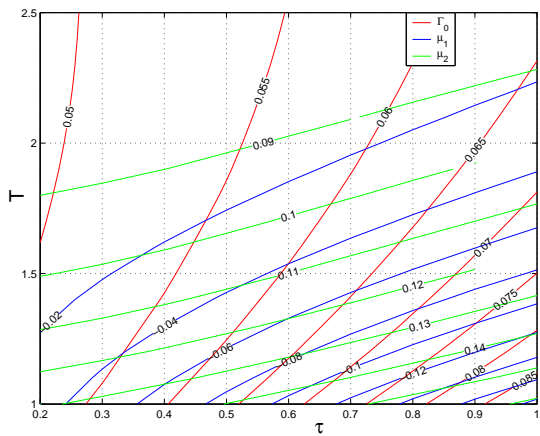}
 \caption{Dependence of $\Gamma_0$ and of $\mu_{1,2}$ on $T,\tau$
 (measured in the units set by re-scaling the pulse width to unity).}
 \label{fig:Gamma_mu} \end{figure}

Intending  to analyze the dependence of the parameters
$\Gamma_0\equiv-(D_\xi Z/I_0)\ln B_0$, $\mu_1$,
and $\mu_2$ on the measurement procedure, we present here the results of our calculations for a
simple model case. We assume the Lorentzian shape of the optical filter: ${\cal K}_f{\bm\varPsi}
=\int_0^\infty \mathrm dt'\exp(-t'/\tau) {\bm \varPsi}(t-t')/\tau$. Then, as it follows from Eq.
(\ref{phiphi}) the statistics of the inhomogeneous contribution, ${\cal K}\bm\phi$, is governed by
the PDF, ${\cal P}$:
 \begin{eqnarray}
 \ln{\cal P}(\bm\phi)=-\frac{1}{D_\xi Z}\int\mathrm
 dt\left[\left|{\cal K}{\bm\phi}\right|^2
 +\tau^2\left|\partial_t{\cal K}{\bm\phi}\right|^2\right].
 \label{Pphi}\end{eqnarray}
The large value of OSNR justifies the saddle-point approximation for calculating $B$. The
saddle-point equation, found by varying Eq. (\ref{Pphi}) over $\phi$, is
 \begin{eqnarray}
 \left[\tau^2\partial_t^2-1
 -uG(t)\right]{\cal K}{\bm\phi}
 =uG(t){\cal K}{\bm\varphi},
 \label{sp1} \end{eqnarray}
where $u$ is a parameter to be extracted from the self-consistency condition (\ref{nnn}). $B$ can
be estimated by ${\cal P}(\bm\phi_0)$, with $\bm\phi_0$ being the solution of Eqs.
(\ref{nnn},\ref{sp1}) for $I=I_d$. Next, we assume that $G(t)=1$ at $|t|<T$ and it is zero
otherwise. Then, for a given value of $u$, the solution of Eq. (\ref{sp1}) can be found
explicitly. The value of the parameter $u$ is fixed implicitly by Eq. (\ref{nnn}). Thus $u$ (and
then $B$) can be found perturbatively in $h_j$, i.e. as $u\approx u_0+\delta u$, $\delta u\ll
u_0$, where $u_0$ is the solution of the system (\ref{nnn},\ref{sp1}) at $h_j=0$. For the Gaussian
shape of the initial pulse, $\varPsi_0\propto\exp[-t^2/(2b^2)]$, and $I_d$ being the half of
the ideal output intensity (corresponding to $\varPsi(Z)=\varPsi_0$), the numerically found
dependencies of $\Gamma_0$, $\mu_{1,2}$ on $\tau$ and $T$ (measured in the units of pulse width
$b$) are shown in Fig. \ref{fig:Gamma_mu}.

The PDF ${\cal S}(B)$ of $B$, (which appears as the result of averaging over many birefringent
disorder realizations) can be found by recalculating the statistics of $H_j$ using Eq. (\ref{hh})
followed by substituting the result into the corresponding expression that relates $B$ to $H_j$.
Our prime interest is to describe the PDF tail corresponding to $H_j$ substantially exceeding
their typical value $\sqrt{D_m Z}$, however, $H_j$ remaining much smaller than the signal duration
$b$. In this range one gets the following estimate for the differential probability ${\cal S}(B)\,
\mathrm d B$:
 \begin{equation}
 a)\ \exp\left[-\frac{D_\xi^2Z b^2}{2D_m\mu_1^2I_0^2}
 \ln^2\left(\frac{B}{B_0}\right)\right]
 \frac{dB}{B}, \quad b)\
 \frac{B_0^\alpha\,\mathrm d B}{B^{1+\alpha}}\,,
 \label{PDFBER} \end{equation}
where (a) corresponds to the no-compensation case, (b) is related to the optimal
``setting the clock" case, and
$\alpha\equiv D_\xi b^2/(2\mu_2D_m I_0)$. Note, that the result corresponding to case (b) shows a
steeper decay compared to case (a), which is a natural consequence of the compensation procedure
applied.

Summarizing, our major result is the emergence of the extremely long tail (\ref{PDFBER}) in the
PDF of BER. Note that Eq. (\ref{PDFBER}) shows a complex ``interplay'' of noise and disorder. To
illustrate this focal point of our analysis consider an example of a fiber line with the parameters
$\Gamma_0=0.06$, $\mu_1=0.06$ and $\mu_2=0.12$, which is also characterized by typical bit error
probability, $B_0=10^{-12}$, corresponding to $I_0/[D_\xi Z]\approx460$. Let us also assume that
the PMD coefficient, $k=\sqrt{12D_m}$, is $0.14$ $ps/\sqrt{km}$, the pulse width, $b$, is $25$
$ps$, and the length of the fiber, $Z$, is $2,500$ $km$, i.e. $D_m Z/b^2\approx 6\cdot 10^{-3}$.
Then we can find a probability for $B$ to exceed, say, $10^{-8}$, that is to become at least $4$
orders larger than $B_0$. Our answers (\ref{PDFBER}a) and (\ref{PDFBER}b) give for this probability
$10^{-4}$ and $10^{-6}$, respectively, which essentially exceed any naive Gaussian or exponential
estimate of the PDF tail. Also note that even though an extensive experimental (laboratory and
field trial) of our analytical result would be of a great value, some numerical results,
consistent with Eq. (\ref{PDFBER}) are already available. Thus, Fig. 2a of \cite{01XSKA} re-plotted
in log-log variables shows the relation between $\ln S$ and $\ln B$ close to the linear one given
by Eq. (\ref{PDFBER}b).

We are thankful to I. Gabitov for numerous helpful discussions. We also thank P. Mamyshev for
valuable remarks. The support of LDRD ER on ``Statistical Physics of Fiber Optics Communications"
at LANL is acknowledged.

\end{document}